\documentclass[12pt]{article}
\pdfoutput=1
\usepackage{epsf,amsfonts,amssymb,epsfig,amsmath,graphics,slashed}
\usepackage{hyperref,graphicx,subfig}
\newcommand{\nn}{\notag \\}

\begin{document}

\makeatletter
\renewcommand{\theequation}{\thesection.\arabic{equation}}
\@addtoreset{equation}{section}
\makeatother

\baselineskip 18pt

\begin{titlepage}

\vfill

\begin{flushright}
Imperial/TP/2011/JG/05\\
\end{flushright}

\vfill

\begin{center}
   \baselineskip=16pt
   {\Large\bf  Spatially modulated instabilities\\ of magnetic black branes}
  \vskip 1.5cm
      Aristomenis Donos, Jerome P. Gauntlett and Christiana Pantelidou\\
   \vskip .6cm
      \begin{small}
      \textit{Blackett Laboratory, 
        Imperial College\\ London, SW7 2AZ, U.K.}
        \end{small}\\*[.6cm]

\end{center}

\vfill

\begin{center}
\textbf{Abstract}
\end{center}

\begin{quote}
We investigate spatially modulated instabilities of magnetically charged
$AdS_{2}\times\mathbb{R}^2$, $AdS_{3}\times\mathbb{R}^2$ and
$AdS_{2}\times\mathbb{R}^3$ backgrounds 
in a broad class of theories, including those arising from KK reductions of ten and eleven dimensional supergravity. We show that magnetically charged black brane 
solutions in $D=4,5$ spacetime dimensions, whose zero temperature near horizon 
limit approach these backgrounds, can have instabilities that are dual to phases with
current density waves that spontaneously break translation symmetry. Our examples include
spatially modulated instabilities for a new class of magnetic black brane solutions of $D=5$ $SO(6)$ gauged supergravity, that we construct
in closed form, which are dual
to new phases of $N=4$ SYM theory.
\end{quote}

\vfill

\end{titlepage}
\setcounter{equation}{0}

%%%%%%%%%%%%%%%%%%%%%%%%%%%%%%%%%%%%%%%%%%%%%%%%%%%%%%%%%%%%%%%%%%%%%%%
%\tableofcontents
%%%%%%%%%%%%%%%%%%%%%%%%%%%%%

%\begin{thebibliography}{99}

\section{Introduction}

The AdS/CFT correspondence can be used to analyse the
dynamics of strongly coupled gauge theories at finite temperature by constructing
and studying black hole solutions of $D=10$ and $D=11$ supergravity. 
The stability properties of the black hole solutions play an important role as they
are related to the thermodynamical stability properties of the gauge theories. Typically, at the onset of an instability
new black hole solutions appear which are dual to new phases of the dual gauge theory, which may have
applications to condensed matter systems.

Several new classes of black hole solutions have been found in this way, principally in the context of
{\it electrically} charged black brane solutions of Einstein-Maxwell theory. Recall that these are
simply AdS-Reissner-Nordstr\"om (AdS-RN) black branes and are dual to field theories at finite
temperature and charge density.
A prominent class of instabilities appear after embedding the solutions into theories of gravity with
additional charged fields. The resulting instabilities lead to black brane solutions
with charged hair that spontaneously break a global abelian symmetry and hence are 
dual to superfluid phases. Such superfluid black branes were first analysed in phenomenological
theories of gravity \cite{Gubser:2008px,Hartnoll:2008vx,Hartnoll:2008kx}
and then in consistent truncations of $D=10,11$ supergravity 
\cite{Denef:2009tp,Gauntlett:2009dn,Gauntlett:2009bh,Gubser:2009qm}.  They were first
studied using D-brane probes in \cite{Ammon:2008fc}.

The electrically charged AdS-RN black branes can also have spatially modulated instabilities 
leading to interesting new classes of black brane solutions that are dual to phases which
spontaneously break translation invariance.
Such instabilities have been investigated for a class of $D=5$
gravity theories with a single gauge-field and a Chern-Simons coupling 
in \cite{Nakamura:2009tf,Ooguri:2010kt}
(for earlier related work see \cite{Domokos:2007kt}). It has also been shown \cite{Ooguri:2010xs} 
that these instabilities are present in the Sakai-Sugimoto
model. A study of the effect of some higher derivative corrections was made in \cite{Takeuchi:2011uk}.

Similar instabilities are also present in $D=4$ \cite{Donos:2011bh}. 
In this case they are associated with  ``striped" black brane solutions which
are dual to phases with both current density waves and charged density waves.
The $D=4$ theories studied in \cite{Donos:2011bh} have a neutral pseudo-scalar field $\varphi$ coupled to
one or two vector fields and are of a form that is very natural in the context of $N=2$ gauged supergravity.
The key couplings in the Lagrangian that drive the instabilities are 
either $\varphi F\wedge F$ or
$\varphi F\wedge G$, where $F,G$ are the field strengths of the two vector-fields.
Indeed these terms give rise to a mixing of the linearised modes via a linear dependence
on the spatial momentum on the black brane.  
It was also shown in \cite{Donos:2011bh} how some of the $D=4$ 
models which exhibit the spatially modulated instabilities can be embedded into $D=10,11$ supergravity. 
Holographic striped instabilities were also studied in the context of probe-branes in
\cite{Bergman:2011rf} utilising, essentially, the same mechanism.

In this paper we show that {\it magnetically} charged black brane solutions 
can also have spatially modulated instabilities. We will analyse several different
models, including some top-down examples. We will first analyse black brane solutions
of Einstein-Maxwell theory.
Recall that in $D=4$ the magnetically charged black brane solutions are again
the standard AdS-RN black brane solutions, while in $D=5$ they have only
been constructed numerically \cite{D'Hoker:2009mm}. 
In both cases, at zero temperature, the solutions interpolate between 
$AdS_{D}$ in the UV and $AdS_{D-2}\times \mathbb{R}^2$ in the IR. It is natural to
view the $AdS_{D-2}$ region as describing the strongly coupled dynamics of
the lowest Landau-level excitations of the $D-1$ dimensional dual gauge-theory.

We will show that spatially modulated instabilities of the magnetic black branes of Einstein-Maxwell theory
can appear
after embedding them in a class of $D$-dimensional theories of gravity that 
involve two vector fields and a single 
scalar field $\phi$. In these models the key coupling in the Lagrangian that drives 
the instability is now $\phi *F\wedge G$.
The class of $D$-dimensional theories that we consider naturally arise in $N=2$ gauged-supergravity,
and we discuss some specific embeddings into $D=10,11$ supergravity. 
The simplest way to test for instabilities of the finite temperature black brane solutions is to look for
linearised perturbations of the $AdS_{D-2}\times\mathbb{R}^2$ limiting solution
that violate the $AdS_{D-2}$ BF bound. We investigate this in section \ref{sub:instabilities}
and find that spatially modulated instabilities arise very naturally within the class of models that we consider.
We have not yet been able to find examples of these instabilities within any 
consistent truncations of $D=10,11$ supergravity, but we expect that they can be found.

For $D=4$, where the magnetic black brane solutions are known analytically,
we go beyond the near horizon $AdS_{D-2}\times\mathbb{R}^2$ region in section \ref{bbrane},
and analyse linearised instabilities in the full geometry.
More specifically we look for spatially modulated zero modes that
appear just prior to the appearance of an instability. 
For some representative models that have
a spatially modulated instability in the $AdS_{D-2}\times\mathbb{R}^2$ region 
we determine the critical temperatures at
which the the zero modes appear. 
We will explain how these instabilities are associated with
phases that have current density waves without having charge density waves (in contrast to the $D=4$ electrically charged
striped black branes of \cite{Donos:2011bh}).

For models that do not have a spatially modulated instability in the $AdS_{D-2}\times\mathbb{R}^2$ region, it is
possible that there still can be spatially modulated 
instabilities in the full black brane geometry. We have analysed 
a model arising in the $SU(3)$ invariant sector
of $D=4$ $SO(8)$ gauged supergravity, but our numerical results only allow us
to conclude that if the black brane has such an instability it will be at a very low temperature.

In section 5 we discuss instabilities of a different class of magnetic black brane solutions.
They arise in theories of gravity coupled to a scalar field and a single gauge field but cannot
be truncated to Einstein-Maxwell theory. We focus on models that have
$AdS_{D-2}\times\mathbb{R}^2$ solutions and discuss the spatially
modulated instabilities.

An interesting class of $AdS_3\times\mathbb{R}^2$ and $AdS_2\times\mathbb{R}^3$ solutions of $D=5$ $SO(6)$ gauge supergravity, and hence
type IIB supergravity, have recently been discussed in 
\cite{Almuhairi:2010rb,Almuhairi:2011ws}, which carry magnetic charges with respect
to $U(1)^3\subset SO(6)$. These include both supersymmetric and non-supersymmetric examples.
For the former class, an investigation of some instabilities, including
those of the type discussed in \cite{Ammon:2011je}, was made. 
In section 6 we will first present a new magnetic black brane solution in closed form
that at zero temperature and in the near horizon limit approaches the
$AdS_2\times \mathbb{R}^3$ solution of \cite{Almuhairi:2010rb}. By considering
perturbations about the $AdS_2\times \mathbb{R}^3$ solution we
find spatially modulated modes that violate the $AdS_2$ BF bound. This example thus provides a top-down setting
in which magnetic black branes exhibit spatially modulated instabilities. Interestingly it corresponds to a phase of $N=4$ SYM
with both current density waves and charged density waves.

\section{Models extending Einstein-Maxwell}\label{sec2}
We start with Einstein-Maxwell theory in $D$ spacetime dimensions with Lagrangian given by 
\begin{align}\label{eq:LagraEM}
\mathcal{L}= &\left[\tfrac{1}{2}R+\lambda^{2}\,\left(D-3\right)\right]\,\ast 1-\tfrac{1}{2}\ast F\wedge F\, ,
\end{align}
where $F\equiv dA$. The negative cosmological constant has been written in terms of the constant $\lambda$
for convenience.
We are interested in asymptotically $AdS_D$ black brane solutions of this theory that are supported by magnetic flux in a single
$\mathbb{R}^2$-plane. When $D=4$ these solutions are the standard magnetically charged 
AdS-Reissner-Nordstr\"om black brane solutions. 
When $D=5$ the solutions have been constructed numerically in
\cite{D'Hoker:2009mm}. The solutions\footnote{When $D\ge 6$ one can also have magnetic fields
switched on in additional planes. The special case when the skew eigenvalues of the two-form field strength are
all equal was discussed in \cite{D'Hoker:2009mm} and the solutions are
similar to the $D=4,5$ cases depending on whether $D$ is even or odd, respectively.}
have not yet been constructed for $D\ge 6$.

In both the $D=4$ and $D=5$ cases, at zero temperature, the near horizon limit of these black brane
solutions approach
the magnetically charged $AdS_{D-2}\times\mathbb{R}^2$ solutions of \eqref{eq:LagraEM} given by
\begin{align}\label{eq:AdS_solution}
ds^{2}&={L^{2}}\,ds^{2}\left(AdS_{D-2}\right)+dx_{1}^{2}+dx_{2}^{2},\qquad \qquad L^2=\frac{D-3}{2\lambda^2}\nn
F&=\sqrt{2}\lambda\,dx_{1}\wedge dx_{2}\, ,
\end{align}
where $ds^{2}\left(AdS_{D-2}\right)$ is the metric on a unit radius $AdS_{D-2}$ space. We expect that
this will similarly be true in $D\ge 6$.

We can look for instabilities of the magnetically charged black brane solutions by analysing linearised
perturbations about the $AdS_{D-2}\times\mathbb{R}^2$ solutions.  If the perturbations have mass squared
that violate the $AdS_{D-2}$ BF bound, given by
\begin{align}\label{BFbound}
L^2M^2\ge-\frac{(D-3)^2}{4},
\end{align}
then we can conclude that the black brane solution is also unstable. In order to establish the precise 
temperature at which the instability appears one needs to analyse the full finite temperature black brane
solution which, as we mentioned above, is only known in $D=4,5$. We shall return to this point in
section \ref{bbrane}.

We now embed these solutions in a theory of gravity that has 
an additional scalar field, $\phi$, and a massive vector field, $B$,
with Lagrangian
\begin{align}\label{eq:Lagra}
\mathcal{L}= &\tfrac{1}{2}R\, \ast 1-V\left(\phi\right)\,\ast 1-\tfrac{1}{2}\,\ast d\phi\wedge d\phi-\tfrac{1}{2}t\left(\phi\right)\,\ast F\wedge F\nn&-\tfrac{1}{2} v(\phi)\ast G\wedge G-\tfrac{1}{2}m_{v}^{2}\,\ast B \wedge B
-u\left(\phi\right)\,\ast F\wedge G\, 
\end{align}
where $G\equiv dB$ and $m_v^2$ is a constant.
The corresponding equations of motion are given by
\begin{align}\label{eoms}
&R_{\mu\nu}=\tfrac{2}{D-2}V\,g_{\mu\nu}+m^2_v B_{\mu}B_{\nu}+\partial_{\mu}\phi\,\partial_{\nu}\phi
+t \left(F_{\mu\rho}F_{\nu}{}^{\rho}-\tfrac{1}{2\left(D-2\right)}g_{\mu\nu}F_{\rho\sigma}F^{\rho\sigma} \right)\nn
&\qquad\quad+ v\left(G_{\mu\rho}G_{\nu}{}^{\rho}-\tfrac{1}{2\left(D-2\right)}g_{\mu\nu}G_{\rho\sigma}G^{\rho\sigma}\right)
+2u \left( G_{(\mu}{}^\rho F_{\nu)\rho}{}-\tfrac{1}{2\left(D-2\right)}g_{\mu\nu}G_{\rho\sigma}F^{\rho\sigma} \right)\, ,\nn
&d\ast\left(t\, F+u\,G\right)=0\, ,\nn
&d\ast\left(vG+u\,F \right)-\left(-1\right)^{D}\,m_{v}^{2}\,\ast B=0\, ,\nn
&\left(-1\right)^{D}\,d\ast d\phi+V^{\prime}\ast 1+\tfrac{1}{2}t^{\prime}\,\ast F\wedge F
+\tfrac{1}{2}v^{\prime}\,\ast G\wedge G+u^\prime\,\ast F\wedge G=0\, .
\end{align}
We will assume that the functions $V,t,u$ and $v$ have the following expansion
\begin{align}\label{conds}
V(\phi)&=-\lambda^{2}\,\left(D-3\right)+\tfrac{1}{2}\,m_{s}^{2}\,\phi^{2}+\cdots,\nn
t(\phi)&=1-n\,\phi^{2}+\cdots,\nn
u(\phi)&=s\,\phi+\cdots\, , \nn
v(\phi)&=1+\cdots .
\end{align}
where $\lambda,m_s,n$ and $s$ are constant. It is then
consistent to set $\phi=B=0$ in
the equations of motion to recover the equations of motion of the Einstein-Maxwell theory
\eqref{eq:LagraEM}. In particular, both the black brane solutions (for $D=4,5$) and the
$AdS_{D-2}\times\mathbb{R}^2$ solutions of Einstein-Maxwell theory are solutions of this more general class
of theories.

The Lagrangian \eqref{eq:Lagra} has been chosen to provide
a simple setting to display spatially modulated instabilities. There are certainly many ways in
which additional fields and interactions can be incorporated which will lead to 
generalisations of the instabilities that we describe. Note that we will only be considering configurations
for which $F\wedge F=F\wedge G=G\wedge G=0$ and hence various terms one might consider, such as Chern-Simons terms in odd dimensions and coupling to pseudo-scalars in even dimensions, will not play a role.

The form of \eqref{eq:Lagra} for $D=4$ is also rather
natural from the point of $N=2$ gauged supergravity. Indeed we can make contact with a
top-down construction this way. Recall that
$D=4$ $SO(8)$ gauge-supergravity is a consistent truncation of
$D=11$ supergravity. We then consider the further consistent truncation
to the $SU(3)\subset SO(8)$ invariant sector which is described by an $N=2$ $D=4$ supergravity theory coupled
to a vector multiplet and a hypermultiplet \cite{Warner:1983vz,Bobev:2010ib}. Restricting to configurations in which
$F\wedge F=F\wedge G=G\wedge G=0$, where $F$ and $G$ are the field strengths of the two-vector fields,
it is consistent to set to zero the four scalar fields in the hypermultiplet as well
as the imaginary part of the complex scalar field in the vector multiplet. This 
leads\footnote{ e.g. set $\rho=\chi=0$ in eq. (2.22) of \cite{Donos:2011ut}.}
to
a theory with two gauge fields and a real scalar field and we obtain 
equations of motion as in \eqref{eoms}, \eqref{conds} with $m^2_s=-4$, $m^2_v=0$, $\lambda^2=6$,
$n=-1$ and $s=-\sqrt{2}$. This can also be obtained from the semi-consistent $U(1)^4\subset SO(8)$ truncation of
\cite{Chong:2004ce} and one should be aware that the details will affect
the higher order terms in \eqref{conds}. For example, the
sub-truncation of \cite{Chong:2004ce} considered in section 2.2.2 of \cite{Donos:2011ut} has a
$\mathbb{Z}_2$ symmetry which flips the sign of both $\phi$ and $B$ but this symmetry is absent in
the $SU(3)\subset SO(8)$ invariant case.

In $D=5$ the model \eqref{eq:Lagra} also arises in Roman's $N=4^+$ $SU(2)\times U(1)$ gauged
supergravity theory \cite{Romans:1985ps}, 
which is a consistent truncation
of type IIB \cite{Lu:1999bw} 
and $D=11$ supergravity \cite{Gauntlett:2007sm}. We will use the action for Roman's theory given
in eq. (2.14) of \cite{Gauntlett:2007sm}. We set $m=1$, $C=F^1=F^2=0$ and restrict to configurations in
which the Chern-Simons terms can be set to zero. We then define $X=e^{\frac{1}{\sqrt{3}}\phi}$ and also $G\rightarrow \sqrt{\frac{2}{3}}\left(F-\sqrt{2}\,G \right)$, $F^{3}\rightarrow \sqrt{\frac{2}{3}}\left(\sqrt{2}\,F+G \right)$ to obtain
equations of motion as in \eqref{eoms}, \eqref{conds} with
 $n=-4/3$, $m_{s}^{2}=-4$, $m_{v}^{2}=0$, $\lambda=\sqrt{3}$ and $s=-2\sqrt{\frac{2}{3}}$.
It is interesting to point out that starting with the consistent truncation of $SO(6)$ gauged supergravity given
in \cite{Bobev:2010de}, we obtain a model with the same parameters.
Indeed, starting with
the action in eq (2.7) of \cite{Bobev:2010de}, which is only valid for configurations in which the Chern-Simons terms
play no role, one can set $\varphi_i=0$ and either $\beta=0$, $A^1=A^2$ or $\beta=3\alpha$, $A^1=A^3$.

\section{Instabilities of magnetic $AdS_{D-2}\times \mathbb{R}^{2}$}\label{sub:instabilities}
In the models extending Einstein-Maxwell theory with Lagrangian \eqref{eq:Lagra},
we consider the following simple perturbation about the 
$AdS_{D-2}\times \mathbb{R}^{2}$ solution \eqref{eq:AdS_solution}:
\begin{align}\label{fbpert}
\phi=\delta\phi\left(x^{\alpha}\right)\,\cos\left(kx_{1}\right),\qquad 
B=\delta B\left(x^{\alpha}\right)\,\sin\left(kx_{1}\right)\,dx_{2}\, ,
\end{align}
where $x^\alpha$ are coordinates on $AdS_{D-2}$ and  
$k$ is a constant\footnote{Note that the background has $F\wedge F=0$ and 
that the perturbation satisfies $F\wedge G=G\wedge G=0$.}. Introducing the vector ${\bf v}=(\delta\phi,\delta B)$ we find that,
at linearised order,
the equations of motion \eqref{eoms}  
expanded around the solution \eqref{eq:AdS_solution} yield
\begin{align}
\Box_{D-2}{\bf v}-L^2M^2{\bf v}&=0\, ,
\end{align}
where $\Box_{D-2}$ is the Laplacian of the unit radius $AdS_{D-2}$
and the mass matrix is given by 
\begin{align}\label{eq:mass_matrix}
%L^{2}
M^{2}=\left(
   \begin{matrix} 
      \tilde{m}_{s}^{2}+k^{2} & \sqrt{2}\lambda s\,k \\
     \sqrt{2}\lambda s\,k & m_{v}^{2}+k^{2} \\
   \end{matrix}
   \right)\, ,
\end{align}
with $\tilde{m}_{s}^{2}\equiv m_{s}^{2}-2\lambda^2 n$. 
Let us first consider no spatial modulation i.e. $k=0$. We see that just
as in the $D=4$, purely electric case studied in \cite{Donos:2011bh}, if $n$ is positive and
large enough, the BF bound \eqref{BFbound} can be violated leading to
an instability. We have not been able to find any top-down examples where this occurs.

We now consider $k\ne 0$.
The eigenvalues of the mass matrix \eqref{eq:mass_matrix} are given by
\begin{equation}
%L^2 
m_{\pm}^{2}=k^{2}+\frac{1}{2}\,\left(\tilde{m}_{s}^{2}+m_{v}^{2}\right)\pm \frac{1}{2}\,\sqrt{\left(\tilde{m}_{s}^{2}-m_{v}^{2}\right)^{2}+8k^{2}s^{2}\lambda^{2}}\, .
\end{equation}
%Clearly when $\lambda s k$ is large enough the BF bound \eqref{BFbound} 
%will be violated. I
We deduce that if
%In fact if we assume that $\m_v^2$ and $\tilde m_s^2$ do not violate
%the BF biund
\begin{equation}
2s^{2}\lambda^{2}>\left|\tilde{m}_{s}^{2}-m_{v}^{2}\right|\, ,
\end{equation}
the branch $m_{-}^{2}$ develops a minimum at
\begin{equation}\label{minkval}
k_{min}=\frac{1}{2\sqrt{2}\,s\lambda}\,\sqrt{4s^{4}\lambda^{4}-\left(\tilde{m}_{s}^{2}-m_{v}^{2}\right)^{2}}\, ,
\end{equation}
with
\begin{equation}\label{eq:min_ads_mass}
%L^2
m_{min}^{2}=\frac{1}{2}\,\left(\tilde{m}_{s}^{2}+m_{v}^{2}\right)-\frac{1}{8s^{2}\lambda^{2}}\,\left(\tilde{m}_{s}^{2}-m_{v}^{2}\right)^{2} -\frac{1}{2}s^{2}\lambda^{2}\, ,
\end{equation}
and it is possible for this to violate the $AdS_{D-2}$ BF bound \eqref{BFbound}.

We now investigate whether these instabilities of $AdS_{D-2}\times \mathbb{R}^2$ are present
in the top-down models we discussed above.
First consider the $SU(3)$ invariant sector of $D=4$ $SO(8)$ gauged supergravity which has
$m^2_s=-4$, $m^2_v=0$, $\lambda^2=6$,
$n=-1$ and $s=-\sqrt{2}$. This gives rise to $L^2m^2_{min}=-2/9$ (at $k\ne 0)$
which is very close to but does not violate the 
$AdS_2$ BF bound
of $-1/4$.
For the $D=5$ Romans theory we have
 $n=-4/3$, $m_{s}^{2}=-4$, $m_{v}^{2}=0$, $\lambda=\sqrt{3}$ and $s=-2\sqrt{\frac{2}{3}}$.
We find $L^2m^2_{min}=-3/4$ (at $k\ne 0$) which again does not violate the
$AdS_3$ BF bound of $-1$.
In the appendix we will discuss some other top down constructions which are similar to 
\eqref{eq:Lagra} but involve a second scalar field; we find that they also
do not lead to a violation of the $AdS_{D-2}$ BF bound \eqref{BFbound}. 

\section{Instabilities of $D=4$ magnetic AdS-RN black branes}\label{bbrane}
In this section we analyse spatially modulated instabilities of the full magnetic black brane solutions in the models of section 2.
More precisely, we look for zero modes that appear at the onset of instabilities.
Since the $D=5$ magnetic black brane solutions are only known
numerically, we restrict our considerations to $D=4$.

Fixing the cosmological constant by setting $\lambda^{2}=6$  the 
$D=4$ magnetic black brane solution of Einstein-Maxwell theory is the AdS-RN solution
given by
\begin{align}\label{eq:RNsol}
ds^{2}&=-f\,dt^{2}+\frac{dr^{2}}{f}+r^{2}\,\left(dx^{2}+dy^{2}\right)\, ,\nn
F&=r_{+}\,dx\wedge dy\, ,
\end{align}
where
\begin{align}
f&=2r^{2}-\left(2r_{+}^{2}+\frac{1}{2}\right)\frac{r_{+}}{r}+\frac{1}{2}\frac{r_{+}^{2}}{r^{2}}\, .
\end{align}
The zero temperature limit  of this black hole is obtained when $r_{+}=1/\sqrt{12}$ and we note that
\eqref{eq:AdS_solution} can be recovered as the near horizon limit after rescaling the spatial coordinates by $r_+$.

\subsection{First order perturbations}
For simplicity, we now assume that $m_{s}^{2}=-4$ and $m_{v}^{2}=0$ which
covers the case of $\mathcal{N}=8$ gauged supergravity. To construct the zero modes we
consider the perturbation
\begin{align}\label{eq:bh_perturbations}
 \phi=\delta\phi\left(r\right)\cos\left(kx\right)\, ,\qquad
B=\delta B\left(r\right)\sin\left(kx\right)\,dy\, .
\end{align}
The equations of motion \eqref{eoms} then yield
\begin{align}\label{eq:bh_pert_eom}
\frac{1}{r^{2}}\,\partial_{r}\left(r^{2}f\partial_{r}\delta\phi\right)+\left[4
-\frac{k^{2}}{r^{2}}+\frac{nr_{+}^{2}}{r^{4}}\right]\delta\phi-\frac{skr_{+}}{r^{4}}\delta B&=0\, ,\notag\\
\partial_r\left(f\partial_{r}\delta B\right)-\frac{k^{2}}{r^{2}}\delta B-\frac{skr_{+}}{r^{2}}\delta\phi&=0\, .
\end{align}
At the horizon $r=r_{+}$ we impose the
boundary conditions
\begin{align}\label{eq:nh_expansion}
\delta\phi=\phi_{0}+{\cal O}\left(r-r_{+}\right),\qquad
\delta B=b_{0}+{\cal O}\left(r-r_{+}\right)\, .
\end{align}
Since we are dealing with a linear and homogeneous system of equations, we can use a scaling symmetry to set $\phi_{0}=1$. At infinity, on the other hand, we have the asymptotic expansion
\begin{align}
\delta\phi=\frac{\phi_1}{r}+\frac{\phi_2}{r^2}+\cdots,\qquad
\delta B=b_0+\frac{b_{1}}{r}+\cdots\, .
\end{align}
We are only interested in zero modes that spontaneously break translation invariance so we shall
demand $b_0=0$. Note that since we have chosen $m^2_v=0$, the massless gauge-field
is dual to a conserved current $j^B$ in the dual field theory.
For definiteness, we will assume that $\phi$ is dual to an operator ${\cal O}_\phi$ with
dimension $\Delta=1$, as it is in $SO(8)$ gauged supergravity when we quantise with maximal supersymmetry, 
and thus we will also
demand that $\phi_2=0$.

\begin{figure}
\centering
\subfloat[$n=-1$, various $s$]{\includegraphics[width=7cm]{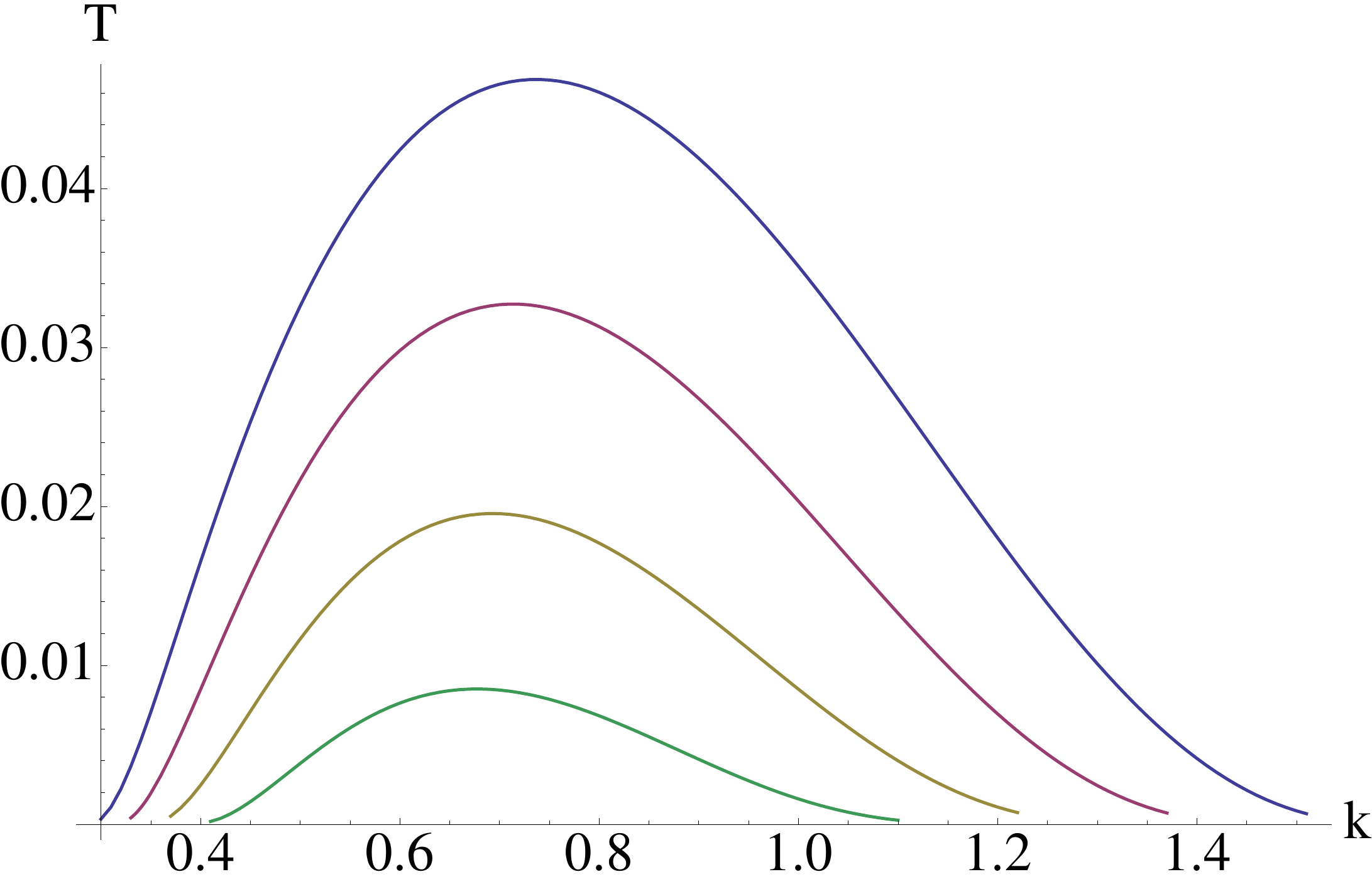}\label{fig:a}}
\subfloat[$s=\sqrt{2}$, various $n$]{\includegraphics[width=7cm]{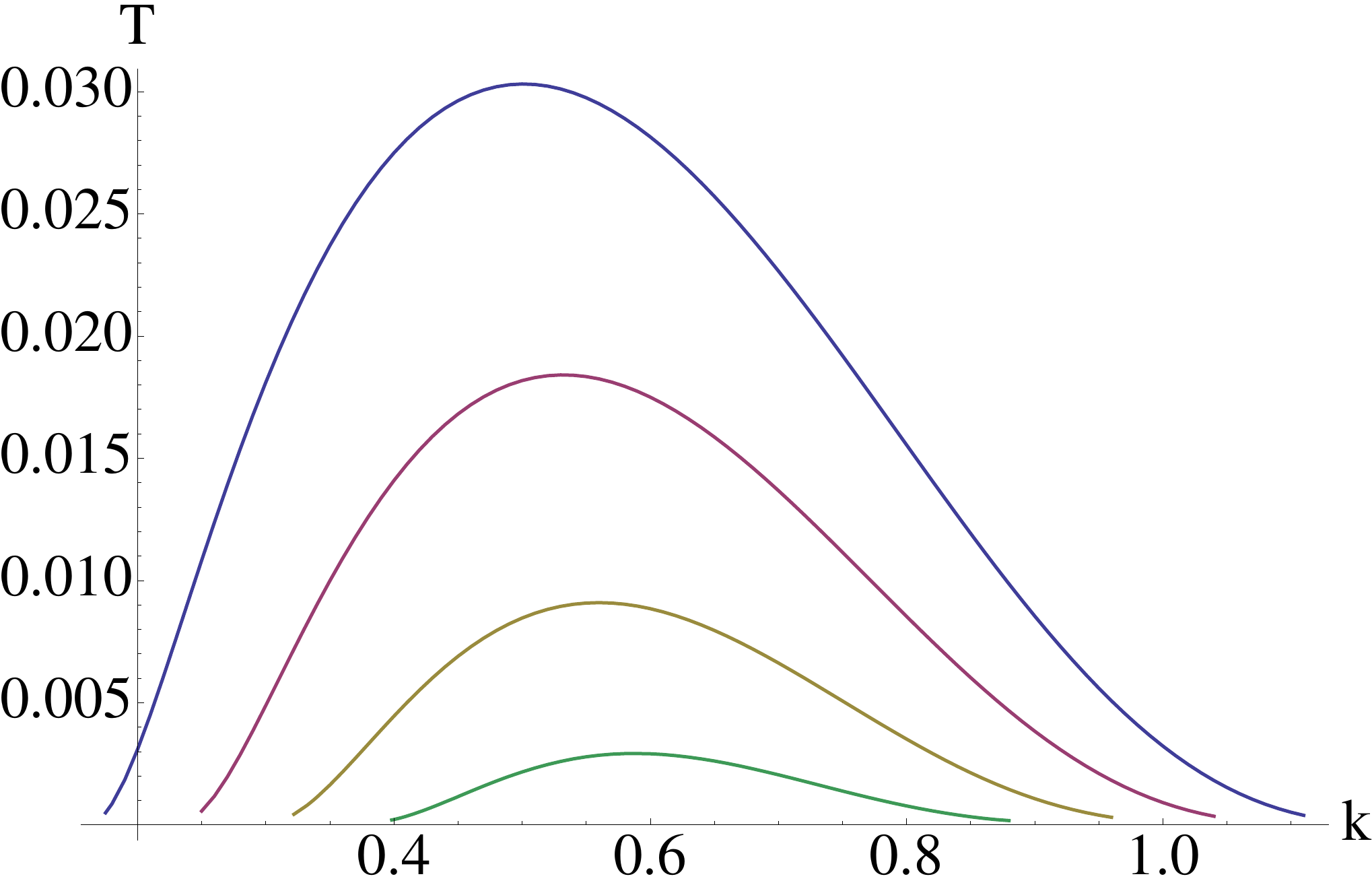}\label{fig:b}}
\caption{Plots of critical temperatures $T$ versus $k$ for the existence of normalisable zero modes
about the $D=4$ magnetically charged
AdS-RN black brane solutions in the models of section \ref{sec2}. All cases have $m^2_s=-4$ and $m_{v}^{2}=0$. 
The left frame has $n=-1$ and, from the top going down, $s= 2,1.9,1.8,1.7$. The right frame
has $s=\sqrt{2}$ and, from the top going down, $n= -0.25,-0.375,-0.5,-0.675$. Note that ${\cal N}=8$ gauged supergravity has $n=-1$ and $s=\sqrt {2}$.}
\label{fig:charged_VEVs_single}
\end{figure}

We have solved numerically the differential equations with these boundary conditions
for two particular models, and determined the critical temperature as a function of $k$
at which the spatially modulated zero modes appear.  We have displayed some of our results in
figure \ref{fig:charged_VEVs_single}.
The first frame has
$n=-1$ and various values of $s= 2,1.9,1.8,1.7$ and the second frame 
has $s=\sqrt{2}$ and various values of $n= -2/8,-3/8,-4/8,-5/8$. All of these
cases have a violation of the $AdS_2$ BF bound in the IR at finite values of $k$.
Note that the values $n=-1$ and $s=-\sqrt{2}$ are relevant for
the magnetic black brane embedded in the $SU(3)$ invariant sector of $D=4$ $\mathcal{N}=8$ gauged 
supergravity. Unfortunately we have not been able to stabilise our numerics for
these values. All that we can conclude is that if there is an instability for these
values of $n,s$ it will be at very low temperatures indeed, as indicated by the figures.

For a given model, at the highest critical temperature $T_c$ at which a static normalisable
zero mode appears (the maxima of the curves in figure \ref{fig:charged_VEVs_single}),
a new branch of black brane solutions 
appear. This new branch will have a spatial modulation fixed by $k_c$, where 
$k_c$ is the critical wave-number corresponding to $T_c$.
These black branes, assuming that they are thermodynamically preferred, describe
a spatially modulated phase in the dual field theory in which, near $T_c$,
\begin{align}
\langle
{\cal O_\phi}
\rangle\sim \cos k_c x\, ,\qquad
\langle j^B_y\rangle \sim \sin k_c x\, ,
\end{align}
where ${\cal O_\phi}$ is the operator dual to $\phi$ and $ j^B$ is the current dual to $B$.
In particular, there is a current density wave. In \cite{Donos:2011bh}, spatially modulated
instabilities of electrically charged black-branes were studied. At first order, current density waves
were observed and then at second order, charged density waves. We now show that this does not
occur for the magnetically charged black branes being considered here.

\subsection{Second order perturbations}
We now consider the second order perturbations. We find that a consistent set of
equations is obtained  if we take
\begin{align}\label{eq:2nd_expansion}
\phi&=\epsilon\left[ \delta\phi\left(r\right)\cos\left(kx\right)\right]+  \epsilon^{2}\left[\phi^{(0)}(r)+\phi^{(1)}(r)\,\cos\left(2kx\right)\right]\, ,\nn
B_{y}&=\epsilon \left[\delta B\left(r\right)\sin\left(kx\right)\right]+  \epsilon^{2}\left[b_{y}^{(1)}(r)\,\sin\left(2kx\right)\right]\, ,\nn
\delta g_{tt}&=\epsilon^{2} \left[ h_{tt}^{(0)}\left(r\right)+h_{tt}^{(1)}\left(r\right)\,\cos\left(2kx \right) \right]\, ,\nn
\delta g_{xx}&=\epsilon^{2} \left[ h_{xx}^{(0)}\left(r\right)+h_{xx}^{(1)}\left(r\right) \,\cos\left(2kx\right)\right]\, ,\nn
\delta g_{yy}&=\epsilon^{2} \left[ h_{yy}^{(0)}\left(r\right)+h_{yy}^{(1)}\left(r\right) \,\cos\left(2kx\right)\right]\, ,\nn
\delta A_{y}&=\epsilon^{2}\left[a_{y}^{(1)}(r)\,\sin\left(2kx\right)\right]\, ,
\end{align}
where $\epsilon$ is a small parameter that can be taken to be 
$\epsilon^2=(T-T_c)/T_c$. 
Expanding the equations of motion \eqref{eoms} to $O\left(\epsilon^{2}\right)$ we obtain a set of ordinary differential equations. The functions $\phi^{(\alpha)}$, $b_{y}^{(\alpha)}$, $h_{xx}^{(\alpha)}$, $h_{yy}^{(\alpha)}$ and 
$a_{y}^{(\alpha)}$ satisfy a system of  inhomogeneous second order equations, the function $h_{tt}^{(0)}$ satisfies a first order inhomogeneous equation while an algebraic equation completely determines $h_{tt}^{(1)}$.

Thus, at second order, we see that in the dual field theory the stress tensor is becoming spatially modulated, as expected. Furthermore, the current, $j^A$, dual to the gauge-field $A$, is also becoming spatially modulated. For the branch appearing at $T_c$ with modulation fixed by $k_c$
we have
\begin{align}
\langle j^A_y\rangle \sim \sin 2k_cx\, ,
\end{align}
and hence the current density wave for $j^A$ has half the period of that for $j^B$. 
The absence of $A_t$ and $B_t$ terms in \eqref{eq:2nd_expansion} implies that there
are no CDWs as commented above.

\subsection{The non-linear ansatz}
We can also investigate
the structure of the spatially modulated black brane solutions by 
finding a consistent ansatz that contain the solutions. Concretely, we consider
\begin{align}\label{eq:non_lin_ansatz}
ds^{2}=&-e^{2\alpha \left(r,x\right)}\,dt^{2}+dr^{2}+e^{2\beta_{1}\left(r,x\right)}\,dx^{2}+e^{2\beta_{2}\left(r,x\right)}\,dy^{2}\, ,\nn
A=&a\left(r,x\right)\,dy,\quad B=b\left(r,x\right)\,dy,\quad \phi=\phi\left(r,x\right)\, .
\end{align}
This includes the magnetic black brane background \eqref{eq:RNsol} as well as the perturbations considered 
in \eqref{eq:2nd_expansion} after a simple redefinition of the coordinate $r$. Note, in particular, that
this anstaz is not associated with CDWs.

To see that it is a well defined ansatz, we proceed as follows. We first find that the equations of motion
for the matter fields leads to three pde's, second order in $r$ and $x$, for the three functions
$a\left(r,x\right), b\left(r,x\right),\phi\left(r,x\right)$. We next observe
that if we write the Einstein's equations in the form
$E_{\mu\nu}=0$, then there are five non-trivial components. The equations
$E_{tt}=E_{xx}=E_{yy}=0$ can also be written as three pdes, second order in $r$ and $x$, for the three functions
$\alpha(r,x),\beta_1(r,x),\beta_2(r,x)$. This leaves two more equations  $E_{rr}=E_{rx}=0$.
However, the Bianchi identities give two relations
\begin{align}
&e^{-\beta_1}\partial_{r}\left(e^{\alpha +\beta_{1}+\beta_{2}}E_{rx} \right)=
-\partial_{x}\left(e^{\alpha -2\beta_{1}+\beta_{2}}E_{xx}\right)
+\frac{1}{2}e^{\alpha +\beta_{2}}\left(
E_{tt}\,\partial_{x}e^{-2\alpha }-E_{yy}\,\partial_{x}e^{-2\beta_{2}}\right)\, ,\nn
%&e^{-\alpha -\beta_{2}}\partial_{x}\left(e^{\alpha -2\beta_{1}+\beta_{2}}E_{xx}\right)+e^{-\alpha -\beta_{1}-\beta_{2}}\,\partial_{r}\left(e^{\alpha +\beta_{1}+\beta_{2}}E_{rx} \right)=\notag\\
%&\qquad \frac{1}{2}E_{tt}\,\partial_{x}e^{-2\alpha }-\frac{1}{2}E_{yy}\,\partial_{x}e^{-2\beta_{2}}\\
&\partial_{r}\left(e^{\alpha +\beta_{1}+\beta_{2}}E_{rr} \right)=
-\partial_{x}\left(e^{\alpha -\beta_{1}+\beta_{2}}E_{rx}\right)\nn
&\qquad\qquad\qquad\qquad+\frac{1}{2}e^{\alpha +\beta_1+\beta_{2}}\left(
E_{tt}\,\partial_{r}e^{-2\alpha }-E_{yy}\,\partial_{r}e^{-2\beta_{2}} -E_{xx}\,\partial_{r}e^{-2\beta_{1}}\right)\, .
%&e^{-\alpha -\beta_{1}-\beta_{2}}\,\partial_{r}\left(e^{\alpha +\beta_{1}+\beta_{2}}\,E_{rr} \right)+e^{-\alpha -\beta_{1}-\beta_{2}}\,\partial_{x}\left(e^{\alpha %-\beta_{1}+\beta_{2}}E_{rx} \right)=\notag\\
%&\qquad \frac{1}{2}E_{tt}\,\partial_{r}e^{-2\alpha }-\frac{1}{2}E_{yy}\,\partial_{r}e^{-2\beta_{2}}
\end{align}
From these we observe that if the three equations
$E_{tt}=E_{xx}=E_{yy}=0$ are satisfied everywhere, then we also have $E_{rr}=E_{rx}=0$ everywhere
provided that we just demand that $E_{rr}=E_{rx}=0$ on a Cauchy surface $r=r_{0}$. Thus, the ansatz leads to a well defined
set of equations, and in particular, will lead to solutions with spatially modulated current density waves without
charge density waves.

\section{Instabilities of $AdS_{D-2}\times\mathbb{R}^2$ in other models}
We now briefly discuss instabilities of magnetic solutions
in models that involve a scalar field coupled to
just a single gauge-field which, unlike the models considered above, cannot
be truncated to Einstein-Maxwell theory. 
We consider the Lagrangian
\begin{align}\label{eq:Lagra2}
\mathcal{L}= &\tfrac{1}{2}R\, \ast 1-V\left(\phi\right)\,\ast 1-\tfrac{1}{2}\,\ast d\phi\wedge d\phi-\tfrac{1}{2}t\left(\phi\right)\,\ast F\wedge F
\end{align}
and note that the equations of motion can be obtained from \eqref{eq:Lagra},\eqref{eoms}. We now consider more general $V,t$ than the expansions given in \eqref{conds} and we note, in particular, that if $V'(0)$ or $t'(0)$ are non-zero then we cannot consistently set $\phi=0$ in the equations of
motion. Particular examples of these models have been considered in the context of AdS/CMT from a different point of view
in \cite{Goldstein:2009cv,Charmousis:2010zz,Gouteraux:2011ce}.

What is of interest here is that for certain $V,t$ these models can 
have magnetically charged asymptotically $AdS_D$ black branes that approach
$AdS_{D-2}\times\mathbb{R}^2$ in the IR at zero temperature. It would go beyond
the scope of this paper to make a detailed analysis of these magnetic black brane solutions.
Instead we will focus on the $AdS_{D-2}\times\mathbb{R}^2$ solutions and see how
spatially modulated instabilities can show up as perturbations that 
have imaginary $AdS_{D-2}$ scaling dimensions. 

We first note that the equations of motion to \eqref{eq:Lagra2} admit the 
magnetically\footnote{An analogous analysis can be carried out for electrically charged $AdS_2\times\mathbb{R}^{D-2}$ solutions.}
charged
$AdS_{D-2}\times\mathbb{R}^2$ solution with constant scalar field, $\phi=\phi_0$, and
\begin{align}
ds^{2}&=L^{2}\,ds^{2}\left(AdS_{D-2}\right)+dx_{1}^{2}+dx_{2}^{2},\qquad\qquad 
L^{2}=-\frac{(D-3)t^{\prime}\left(\phi_{0}\right)}{2V^{\prime}\left(\phi_{0}\right)t\left(\phi_{0}\right)}\, ,\notag\\
F&=\left(-\frac{2\,V^{\prime}\left(\phi_{0} \right)}{t^{\prime}\left(\phi_{0}\right)}\right)^{1/2}\,dx_{1}\wedge dx_{2}\, ,
%L^{2}&=-\frac{t^{\prime}\left(\phi_{0}\right)}{2V^{\prime}\left(\phi_{0}\right)t\left(\phi_{0}\right)}
\end{align}
provided that
\begin{align}
t^{\prime}\left(\phi_{0}\right)V\left(\phi_{0}\right)&=(D-3)V^{\prime}\left(\phi_{0}\right)t\left(\phi_{0}\right)\, ,\notag\\
t^{\prime}\left(\phi_{0}\right)V^{\prime}\left(\phi_{0}\right)&<0\, ,\qquad
t\left(\phi_{0}\right)>0\, .
\end{align}
We have not yet found a top-down embedding of such solutions, but we expect that they can be found.

We next consider spatially modulated instabilities of this solution. Let us write the metric on the unit radius $AdS_{D-2}$ space
as 
\begin{align}\label{adsmet}
ds^2(AdS_{D-2})=\rho^2(-dt^2+dy_a dy_a)+\frac{d\rho^2}{\rho^2}
\end{align}
 Restricting to $D=4$ for
simplicity, we consider the time independent linear perturbation given by\begin{align}
\delta g_{tt}&=\rho^2h_{tt}\left(\rho \right)\,\cos\left(kx_{1}\right),\qquad \delta g_{x_{i}x_{i}}=h_{ii}\left(\rho \right)\,\cos\left(kx_{1}\right)\, ,\notag\\
%\delta g_{x_{i}x_{i}}&=h_{ii}\left(r \right)\,\cos\left(kx_{1}\right)\notag\\
\delta \phi&= w\left(\rho\right)\,\cos\left(kx_{1}\right),\qquad \delta A=a\left(\rho\right)\,\sin\left(kx_{1}\right)dx_{2}\, .
%\delta A&=a\left(r\right)\,\sin\left(kx_{1}\right)dx_{2}
\end{align}
After substituting into the equations of motion for \eqref{eq:Lagra2}, we find that we can solve
an algebraic equation to obtain $h_{tt}$. This then leads to a second
order differential equation for the 4-vector ${\bf v}\equiv (h_{11}, h_{22}, w,a)$ whose coefficients
involve the data $\{t(\phi_0),V(\phi_0),t'(\phi_0),V'(\phi_0),t''(\phi_0),V''(\phi_0)\}$
and $k$.
These equations admit modes of the form ${\bf v}={\bf v}_0\rho^\delta$ where ${\bf v}_0$ is a constant vector and 
$\delta$ is the scaling dimension.
By suitable choice of the data it is simple to obtain complex values for
$\delta$,with $k\ne 0$, which corresponds to a violation of the $AdS_{D-2}$ BF bound.

\section{An instability in type IIB supergravity}\label{topdownsec}
We now consider instabilities of magnetic black brane solutions within $SO(6)$ $D=5$ gauged
supergravity and hence within type IIB supergravity. In fact the solutions will lie within
a consistent truncation of $SO(6)$ gauged supergravity that has two scalar fields
$\phi_1,\phi_2$ and $U(1)^3\subset SO(6)$ gauge fields with Lagrangian \cite{Cvetic:1999xp}
\begin{align}\label{eq:SO6_bh}
\mathcal{L}&=\left(R-V\right)\,\ast 1-\frac{1}{2}\,\sum_{a=1}^{2}\,\ast d\phi_{a}\wedge d\phi_{a}-\frac{1}{2}\,\sum_{i=1}^{3}X_{i}^{-2}\,\ast F^{i}\wedge F^{i}+F^{1}\wedge F^{2}\wedge A^{3}
\end{align}
where $F^i=dA^i$ and
\begin{align}
V&=-4\,\sum_{i=1}^{3}X_{i}^{-1}\notag\\
X_{1}&=e^{-\frac{1}{\sqrt{6}}\phi_{1}-\frac{1}{\sqrt{2}}\phi_{2}},\quad X_{2}=e^{-\frac{1}{\sqrt{6}}\phi_{1}+\frac{1}{\sqrt{2}}\phi_{2}},\quad X_{3}=e^{\frac{2}{\sqrt{6}}\phi_{1}}
\end{align}

Our first result is that this theory admits the following magnetically charged
black hole solutions
\begin{align}\label{bhasol}
ds_{5}^{2}&=-f\,dt^{2}+\frac{dr^{2}}{f}+r^{2}\,\left(dx_{1}^{2}+dx_{2}^{2}+dx_{3}^{2} \right)\notag\\
F^{1}&=\epsilon_1B\,dx_{2}\wedge dx_{3},\quad F^{2}=\epsilon_2B\,dx_{3}\wedge dx_{1},\quad F^{3}=\epsilon_3B\,dx_{1}\wedge dx_{2}
\end{align}
where
\begin{align}
f&=r^{2}-\frac{r_{+}^{4}}{r^{2}}+\frac{B^{2}}{2r^{2}}\,\log{\frac{r_{+}}{r}}
\end{align}
and $\epsilon_i=\pm1$.
We choose $0\le B\le 2\sqrt{2}r_{+}^{2}$ so that
the outer event horizon is located at $r=r_{+}$ and the temperature is $T=\frac{8r_{+}^{4}-B^{2}}{8\pi r_{+}^{3}}$. At zero temperature, when $B=2\sqrt{2}r_{+}^{2}$, the near horizon limit of \eqref{bhasol}
approaches the magnetic $AdS_2\times\mathbb{R}^3$ solution constructed
in \cite{Almuhairi:2010rb}
\begin{align}\label{also}
ds_{5}^{2}&=\frac{1}{8}\,ds^{2}\left(AdS_{2}\right)+dx_{1}^{2}+dx_{2}^{2}+dx_{3}^{2}\notag\\
F^{1}&=\epsilon_12\sqrt{2}\,dx_{2}\wedge dx_{3},\quad F^{2}=\epsilon_22\sqrt{2}\,dx_{3}\wedge dx_{1},\quad F^{3}=\epsilon_32\sqrt{2}\,dx_{1}\wedge dx_{2}
\end{align}
(after scaling $x_i\to x_i/r_+$). Note that we can change the signs of two of the spatial coordinates without changing
the $D=5$ orientation and this would change the signs of two of the three $F^i$. Thus, there are two independent solutions, with
fixed orientation, depending on whether $\epsilon_1\epsilon_2\epsilon_3=\pm1$. In fact both solutions have the same
spatially modulated instability as we now show.

We first consider perturbations about the 
$AdS_2\times\mathbb{R}^3$ solution \eqref{also}. Specifically, using the coordinates for $AdS_2$ as in \eqref{adsmet},
we take
% In order to determine how the dual operators scale we consider the perturbation
\begin{align}\label{eq:perturbation}
\delta g_{x_{1}x_{1}}&=-\delta g_{x_{2}x_{2}}=h(\rho)\,\cos(k\,x_{3})\notag\\
%\delta A^{1}_{3}&=\delta A^{2}_{2}=a(r)\,\sin(k\,x_{1})\notag\\
\delta A^{1}&=\epsilon_1a(\rho)\,\sin(k\,x_{3})dx_2,\qquad \delta A^{2}=\epsilon_2a(\rho)\,\sin(k\,x_{3})dx_1,\nn
\delta A^3&=\epsilon_1\epsilon_2\rho u(\rho)\sin(k x_3)dt\, ,\qquad
\delta \phi_{2}=w(\rho)\,\cos(k\,x_{3})
\end{align}
for which the equations of motion yield
\begin{align}\label{eq:lin_system}
8\,\left(\rho^{2}w^{\prime} \right)^{\prime}-\left(12+k^{2}\right)\,w-8\sqrt{2}\,h+8k\,a&=0\notag\\
8\,\left(\rho^{2}a^{\prime} \right)^{\prime}-k^{2}\,a+4k\,w+2\sqrt{2}k\,h-16\sqrt{2}\,\left(\rho\, u\right)^{\prime}&=0\notag\\
8\,\left(\rho^{2} u^{\prime} \right)^{\prime}-k^{2}\, u-4\sqrt{2}\rho\,a^{\prime}&=0\notag\\
8\,\left(\rho^{2}h^{\prime} \right)^{\prime}-\left(k^{2}+8\right)\,h-8\sqrt{2}\,w+4\sqrt{2}k\,a&=0
\end{align}
Notice that we take into account the mixing of the metric and the scalar even for $k=0$ 
(in contrast to the analysis of \cite{Almuhairi:2010rb}).
We now look for solutions of the form $(w,a,u,h)=\mathbf{v}\,\rho^{\delta}$ with $\mathbf{v}$ a constant vector. The system of equations \eqref{eq:lin_system} then takes the form $\mathbf{M}\,\mathbf{v}=0$ where $\mathbf{M}$ is a $4\times 4$ matrix that depends on $k$ only. Demanding that
non-trivial values of $\mathbf{v}$ exist implies that $\det{\mathbf{M}}=0$ and this equation specifies the possible values of $\delta$ as functions of $k$. In \cite{Almuhairi:2010rb}, where only modes with $k=0$ where considered, it was argued that the $AdS_{2}\times \mathbb{R}^{3}$ background is stable. Here, even after properly
taking into account the mixing with the metric, we still find that for $k=0$ the system is stable. However, for general $k$, a numerical analysis shows that there is a range of $k\neq 0$ for which $\delta$ has a non-zero imaginary part signalling an unstable background.

We now turn our attention to spatially modulated perturbations about the full magnetic
black brane solution \eqref{eq:SO6_bh}. Specifically we consider
\begin{align}\label{eq:perturbation2}
\delta g_{x_{1}x_{1}}&=-\delta g_{x_{2}x_{2}}=r^{2}h(r)\,\cos(k\,x_{3})\notag\\
%\delta A^{1}_{3}&=\delta A^{2}_{2}=a(r)\,\sin(k\,x_{1})\notag\\
\delta A^{1}&=\epsilon_1a(r)\,\sin(k\,x_{3})dx_2,\qquad \delta A^{2}=\epsilon_2a(r)\,\sin(k\,x_{3})dx_1,\nn
\delta A^3&=\epsilon_1\epsilon_2 u(r)\sin(kx_3)dt,\qquad
\delta \phi_{2}=w(r)\,\cos(k\,x_{3})
\end{align}
which leads to the linear system of equations
\begin{align}\label{odesbba}
&r \,\left(r^{3}f\,w^{\prime} \right)^{\prime}-\left(2B^2+k^2 r^2-4 r^4\right) w+2\sqrt{2}B k\, a-\sqrt{2}B^2 h=0\notag\\
&r \left(rf\,a^{\prime} \right)^{\prime}-k^{2}\,a+\sqrt{2}Bk\, w+Bk\, h -rB u'=0\notag\\
& \left(r^3\, u^{\prime} \right)^{\prime}-k^{2}rf^{-1}\,u-2Ba' =0\notag\\
&r \,\left(r^{3}f\,h^{\prime}\right)^{\prime}-r^{2}\left(k^2+8 r^2-4 f-2 r f^{\prime}\right)h+2 B k\, a- \sqrt{2}B^2 w=0
\end{align}
At the black hole horizon we impose the following boundary conditions
\begin{align}
w&=w^{(0)}+w^{(1)}\,\left(r-r_{+}\right)+\cdots\notag\\
a&=a^{(0)}+a^{(1)}\,\left(r-r_{+}\right)+\cdots\notag\\
u&=u^{(1)}\,\left(r-r_{+}\right)+\cdots\notag\\
h&=h^{0}+h^{(1)}\,\left(r-r_{+}\right)+\cdots
\end{align}
As usual we are only interested in spatially modulated zero modes that correspond
to spontaneously breaking of translation invariance. Thus, asymptotically as $r\to\infty$ we impose the boundary conditions
\begin{align}
w=\frac{v_{1}}{r^{2}}+\cdots\notag\\
a=\frac{v_{2}}{r^{2}}+\cdots\notag\\
u=\frac{v_{3}}{r^{2}}+\cdots\notag\\
h=\frac{v_{4}}{r^{4}}+\cdots
\end{align}
with the $v_i$ fixing the expectation values of the corresponding operators in
the dual $N=4$ SYM theory. 

Our analysis of the instabilities of the $AdS_2\times\mathbb{R}^3$ solution implies that there
will be solutions of the ODEs \eqref{odesbba} with these boundary conditions at a specific temperature for a given value of $k$. 
Unfortunately the temperatures are very low and so we have not been able to stabilise the numerics.
The highest critical temperature $T_c$ will occur for a critical wave number $k_c$.
At $T_c$ a new branch of spatially modulated black branes will exist with, at leading order,
\begin{align}
\langle
&{\cal O}_{\phi_2}
\rangle\sim \sin k_c x_3\, ,\nn
&\langle j^1_{x_2}\rangle \sim \sin k_c x_3\, ,\qquad
\langle j^2_{x_1}\rangle \sim \sin k_c x_3\,\qquad \langle j^3_t\rangle \sim \sin k_c x_3\,
\end{align}
where ${\cal O}_{\phi_2}$ is the operator dual to $\phi_2$ and
$j^i$ are the three $U(1)$ currents dual to $A^i$ in $N=4$ SYM theory. Observe that $j^1,j^2$ exhibit 
current density waves, while $j^3$ exhibits charge density waves and that
they are in phase with each other. It would be interesting to explore what happens at
next order in perturbation theory.

\section{Final Comments}\label{fcs}
We have shown that magnetically charged black branes can exhibit spatially modulated instabilities.
In the dual field theory, these correspond to the spontaneous breaking of translation invariance via current
density waves, and in some cases also charge density waves, when the field theory is
placed in a magnetic field. It would be very interesting to go beyond our perturbative
analysis and construct the fully back reacted black brane solutions as well as determining
the zero temperature ground states.

For the class of black brane solutions which have an $AdS_{D-2}\times\mathbb{R}^2$ region 
in the IR at zero temperature, we have not yet been able to find any top-down model that exhibits an instability either in the models extending Einstein-Maxwell theory or those considered in section 5. However, we think it is likely that they can be found.
On the other hand for a new class of black brane solutions of $D=5$ $SO(6)$ 
gauge supergravity with an $AdS_2\times \mathbb{R}^3$ region at zero temperature
that we constructed in section \ref{topdownsec}
we did find spatially modulated instabilities. It would be interesting
to explore this example further and determine, for example, whether or not the instability we found is
the dominant instability within type IIB supergravity.
 
Another direction is to extend our analysis to dyonic black brane solutions,
carrying both electric and magnetic charges. For the $D=4$ case the 
dyonic black brane solutions of
Einstein-Maxwell theory are again
of the AdS-RN form. For the $D=5$ case, dyonic black brane solutions 
have been constructed numerically for a class of gravity theories, including minimal gauged supergravity, in 
\cite{D'Hoker:2009bc,D'Hoker:2010rz,D'Hoker:2010ij}. There are a variety of
ways in which these theories can be coupled to additional fields and it is clear that, again, there 
is a rich spectrum of spatially modulated black brane solutions.

\section*{Acknowledgements}
We would like to thank Nikolay Bobev
for helpful discussions. AD is supported by an EPSRC Postdoctoral Fellowship.
JPG is supported by an EPSRC Senior Fellowship and a Royal Society Wolfson Award. JPG would like to
thank the Aspen Center for Physics for hospitality and he acknowledges the 
support of the National Science Foundation Grant No. 1066293. CP is supported by an I.K.Y. Scholarship.

\appendix
\section{Top down perturbations of magnetic $AdS_{D-2}\times\mathbb{R}^2$}

Here we consider perturbations of magnetic $AdS_{D-2}\times\mathbb{R}^2$ solutions
that appear in some top-down models extending Einstein-Maxwell theory
that are similar to \eqref{eq:Lagra}, but involve a second
scalar field. In all cases we find that the models
do not have spatially modulated instabilities. As in section \ref{bbrane}, 
it is still possible that the magnetic
brane solutions do have them, but we will not investigate this here.

We first consider the consistent truncation of $D=11$ supergravity on a $SE_7$ manifold to a 
$D=4$ $N=2$ gauged supergravity \cite{Gauntlett:2009zw}. It is convenient to use
the Lagrangian given in eq (2.6) of \cite{Gauntlett:2009bh} and we then set $h=\chi=a=0$ to obtain
\begin{align}
\mathcal{L}=&\frac{R}{2}-12 \partial U^{2}-\frac{3}{4}\partial V^{2}-3\partial U\partial V-\frac{1}{8}e^{6U+3V}\,F_{\mu\nu}F^{\mu\nu}-\frac{3}{8}e^{-V-2U}\,H_{\mu\nu}H^{\mu\nu}\nn&-9e^{-12U}\left(B-\epsilon A\right)^{2}
+24e^{-8U-V}-3e^{-10U+V}-9e^{-18U-3V}\, ,
\end{align}
with $F=dA$, $H=dB$ and $\epsilon=\pm1$. Note that this truncation is only valid for configurations which satisfy $F\wedge H=0$. The basic $AdS_4$ vacuum uplifts to an $AdS_4\times SE_7$ solution of $D=11$ supergravity:
when $\epsilon=1$ it is the supersymmetric solution and when $\epsilon=-1$ it is the skew-whiffed solution.
We now redefine $\epsilon A\rightarrow \frac{1}{2\sqrt 2}A-\frac{\sqrt{3}}{2\sqrt 2}B$, $B\rightarrow \frac{1}{2\sqrt 2}A+\frac{1}{2\sqrt{6}}B$ and $U\rightarrow \frac{1}{6}\sqrt{1-\frac{1}{\sqrt{13}}}\,u-\frac{1}{6}\sqrt{1+\frac{1}{\sqrt{13}}}\,v$,
$V\rightarrow \frac{1}{3}\sqrt{4-\frac{10}{\sqrt{13}}}\,u+\frac{1}{3}\sqrt{4+\frac{10}{\sqrt{13}}}\,v$. We can then expand around the $AdS_{2}\times \mathbb{R}^{2}$ solution of the model and find that $u$ and $v$ each combine with $B$
as in \eqref{fbpert} to give a mass matrix as in \eqref{eq:mass_matrix} with
\begin{itemize}
%\item For $u$\\
\item $\tilde{m}_{u}^{2}=8\left(7+\sqrt{13}\right),\quad m_{vec}^{2}=48,\quad s=\sqrt{1+\frac{1}{\sqrt{13}}},\quad\lambda=\sqrt{12}$\, .
%\item For $v$\\
\item $\tilde{m}_{v}^{2}=8\,\left(7-\sqrt{13}\right),\quad m_{vec}^{2}=48,\quad s=\sqrt{1-\frac{1}{\sqrt{13}}},\quad\lambda=\sqrt{12}$\, .
\end{itemize}
In both cases we find that the minimum mass-squared eigenvalue of \eqref{eq:mass_matrix}
is at $k=0$ and does not violate the $AdS_2$ BF bound.

We next consider the consistent truncation of $D=11$ supergravity on $S^4\times H^3$ 
to a $D=4$ $N=2$ gauged supergravity that was derived in  \cite{Donos:2010ax}. Starting with the action given in
eq. (4.14) of  \cite{Donos:2010ax} we set $l=-1$, $\chi=\theta=\xi=a=0$, $T=\delta$. Restricting to configurations in which 
$\tilde H\wedge \tilde H=\tilde H \wedge F=0$
we can also set $\beta=0$. We next redefine the scalars via:
$\phi_{0}\rightarrow -\frac{1}{2}\ln2+\frac{1}{\sqrt{15}}\,\left(\sqrt{5+\sqrt{5}}\,\phi_{0}-\sqrt{5-\sqrt{5}}\,\phi_{1}\right)$,
$\phi_{1}\rightarrow 2\,\left(\frac{1}{\sqrt{5+\sqrt{5}}}\,\phi_{0}+\frac{1}{\sqrt{5-\sqrt{5}}}\,\phi_{1}\right)$
and the vectors via:
$A\rightarrow 2^{1/4}\,\left(A-\sqrt{3}\,B \right)$,
$\tilde B\rightarrow 2^{-3/4}\,\left(A+\frac{1}{\sqrt{3}}\,B \right)$.
Focussing on perturbations about the $AdS_2\times \mathbb{R}^2$ solution we find
\begin{itemize}
\item 
$\tilde{m}_{\phi_{0}}^{2}=3\sqrt{2}+\sqrt{10},\quad m_{v}^{2}=2\sqrt{2},\quad s=-\sqrt{1+\frac{1}{\sqrt{5}}},\quad\lambda=\frac{3^{1/2}}{2^{1/4}}$\, .
\item 
$\tilde{m}_{\phi_{1}}^{2}=3\sqrt{2}-\sqrt{10},\quad m_{v}^{2}=2\sqrt{2},\quad s=\sqrt{1-\frac{1}{\sqrt{5}}},\quad\lambda=\frac{3^{1/2}}{2^{1/4}}$\, .
\end{itemize}
The minimum mass-squared eigenvalue of \eqref{eq:mass_matrix} for $\phi_0$ is at $k\ne 0$, while for $\phi_1$ it is at $k=0$ and neither violates the $AdS_2$ BF bound. 

Finally we consider the consistent truncation of type IIB supergravity on an arbitrary $SE_5$ space
derived in \cite{Maldacena:2008wh}. We start with the action given in eq. (4.21)
of \cite{Maldacena:2008wh}. We then redefine the scalars via:
$u\rightarrow -\frac{1}{5\sqrt{5}}\,\left(u-2\sqrt{6}v\right)$,
$v\rightarrow \frac{1}{5}\sqrt{\frac{2}{15}}\,\left(2\sqrt{6}u+v \right)$ and the vectors via
$\mathcal{A}\rightarrow \sqrt{\frac{2}{3}}\left(A-\sqrt{2}B\right)$,
$\mathbf{A}\rightarrow \sqrt{3} B$. Expanding about the $AdS_3\times\mathbb{R}^2$ solution,
we find 
\begin{itemize}
\item 
$\tilde{m}_{u}^{2}=36, \quad m_{vec}^{2}=24, \quad s=-{\frac{2\sqrt 2}{5}},\quad \lambda=\sqrt{3}$\, .
\item
$\tilde{m}_{v}^{2}=16, \quad m_{vec}^{2}=24, \quad s=\frac{4}{\sqrt{15}},\quad \lambda=\sqrt{3}$\, .
\end{itemize}
The minimum mass-squared eigenvalue of \eqref{eq:mass_matrix} for $u$ is at $k\ne 0$, while for $v$ it is at $k=0$ and neither violates the $AdS_3$ BF bound.

%\bibliographystyle{utphys}
%\bibliography{magnstripes}{}

\providecommand{\href}[2]{#2}\begingroup\raggedright

\end{document}